\documentclass[runningheads]{llncs}

\usepackage{accv}

\usepackage{accvabbrv}

\usepackage{graphicx}
\usepackage{booktabs}

\PassOptionsToPackage{numbers, compress}{natbib}

\usepackage[accsupp]{axessibility}  

\usepackage[pagebackref,breaklinks,colorlinks,citecolor=accvblue]{hyperref}

\usepackage{orcidlink}

\begin{document}

\title{Medical Imaging Complexity and its Effects on GAN Performance} 






\renewcommand{\thefootnote}{}
\author{
William Cagas\thanks{Lead Author.} \and
Chan Ko \and
Blake Hsiao \and
Shryuk Grandhi \and
Rishi Bhattacharya \and
Kevin Zhu\thanks{Senior Author.} \and
Michael Lam$^{**}$
}


\authorrunning{W.~Cagas et al.}

\institute{Algoverse AI Research\\ \email{williamgabriel.cagas@gmail.com, kevin@algoverse.us}}

\maketitle

\begin{abstract}
  The proliferation of machine learning models in diverse clinical applications has led to a growing need for high-fidelity, medical image training data. Such data is often scarce due to cost constraints and privacy concerns. Alleviating this burden, medical image synthesis via generative adversarial networks (GANs) emerged as a powerful method for synthetically generating photo-realistic images based on existing sets of real medical images. However, the exact image set size required to efficiently train such a GAN is unclear. In this work, we experimentally establish benchmarks that measure the relationship between a sample dataset size and the fidelity of the generated images, given the dataset's distribution of image complexities. We analyze statistical metrics based on delentropy, an image complexity measure rooted in Shannon's entropy in information theory. For our pipeline, we conduct experiments with two state-of-the-art GANs, StyleGAN 3 and SPADE-GAN, trained on multiple medical imaging datasets with variable sample sizes. Across both GANs, general performance improved with increasing training set size but suffered with increasing complexity.
\end{abstract}

\section{Introduction}
\label{sec:intro}
Machine learning in healthcare is a rapidly growing field with countless applications \cite{shi2020artificial} including disease diagnosis \cite{rahman2023significance}, clinical treatment \cite{shang2019pre}, drug development \cite{nordon2019separating}, and mental health \cite{graham2019artificial}. The machine learning models driving these advances require the collection of high-quality, annotated medical training data, which persists as an arduous task due to privacy concerns surrounding sensitive patient data \cite{PEZOULAS20242892} and the time-intensive nature of labeling \cite{9324763}. To address these issues, synthetic data—artificially generated information mimicking real-world data—has surfaced as a promising solution \cite{guo2024generative}. 

Currently, generative adversarial networks (GANs) remain one of the leading approaches to synthetic data generation \cite{lu2024machinelearningsyntheticdata}. Since its inception in 2014 \cite{goodfellow2014generative}, GANs have gained increasing attention in the medical research community due to their ability to synthesize medical images \cite{Skandarani2023}. However, achieving results with high fidelity remains a difficult task factoring the lack of medical data and prevalence of smaller datasets in the medical domain. With limited data, a GAN's efficacy is directly affected with consequences including mode collapse, where the generator produces a limited variety of outputs \cite{pan2019recent}, and overfitting, where the GAN replicates training data rather than generalizing from it \cite{webster2019detecting}. 
\begin{figure}
  \centering
  \includegraphics[width=0.75\linewidth]{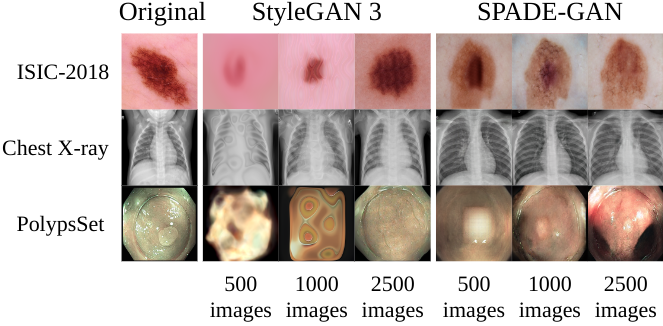}
  \caption{Comparison between original images and synthetic images from StyleGAN 3 and SPADE-GAN based on variable image set sizes.}
  \label{fig:ImageComparisonGraph}
\end{figure}
Various papers such as Wang \etal~'s \cite{wang2018transferring} transfer learning and Robb \etal~'s \cite{robb2020few} Few-Shot GAN (FSGAN) have tried addressing these issues as architecture-centric approaches, achieving increased training efficiency only as a result of the changes in a GAN's structure. However, such approaches are ineffective when making alterations to a GAN's internal structure are not feasible and when time constraints are present. As such, a data-centric approach by providing the GAN with the optimal amount of data to produce high-quality results is more appropriate. Nevertheless, the exact sample set size required to train state-of-the-art GANs is obscure.

In this study, we introduce a data-centric optimization method to create efficient GAN training for medical image synthesis. Our approach investigates how the image complexity distribution of a medical image dataset can be utilized as a measure of training difficulty for a GAN. By doing so, we can ascertain a correlation between the image complexities of the training images and the optimal training set sizes by establishing benchmarks that evaluate the relationship between a sample training set size and the fidelity of the generated images. We hypothesize that given a dataset of a specific image complexity distribution, healthcare professionals can reference the closest image fidelity curve to identify the optimal amount of experimental trials to produce superlative results. Ultimately, our approach can avoid both undertraining and wasteful overtraining by constructing a data-efficient, GAN training pipeline.

\section{Background}
\label{sec:Background}
\paragraph{Generative Adversarial Networks (GANs)} Introduced by Goodfellow \etal~\cite{goodfellow2014generative}, GANs are a class of generative models that consist of two neural networks: a generator $G$, which aims to transform its latent variable distribution $p(z)$ to closely resemble the training data distribution $p(x)$, and a discriminator $D$, which differentiates between the ground truth and data generated by $G$. Training is an adversarial process where $G$ attempts to deceive $D$ into classifying its outputs as real. This two-player minimax game is represented by the following loss function:
\begin{equation}
\min_G \max_D \, V(D, G) = \mathbb{E}_{\mathbf{x} \sim p(\mathbf{x})} \left[ \log D(\mathbf{x}) \right] \mathbb{E}_{\mathbf{z} \sim p(\mathbf{z})} \left[ \log(1 - D(G(\mathbf{z}))) \right].
\end{equation}
Many papers have tried to address data scarcity and computational costs in GAN training architecturally. One proposed approach was transfer learning \cite{wang2018transferring}, which consists of fine-tuning a pre-trained generator and discriminator to the desired domain. However, if the pre-trained models do not align well with the target domain, this could result in even higher data and computational demands \cite{zhuang2020comprehensive}. Another approach, Few-Shot GAN (FSGAN) \cite{robb2020few}, achieved impressive adaptation even with extremely few training examples, albeit at the cost of prolonged training times. This results in the reduced quality and diversity of the synthetic data when time constraints are present.

\paragraph{Image Complexity} Objectively, image complexity can be defined as the variety of features and details within an image. It has been shown that information entropy is a traditional, heuristic-based method of calculating the complexities of images in small-scale datasets \cite{liu2024contrastivelearningimagecomplexity}.

Traditional entropy is a foundational abstraction in information theory introduced by Shannon \cite{6773024}. Used as a measure of uncertainty or “surprise” in data, it is the variation in the distribution of pixel intensities of an image in grayscale format. The equation is defined as
\begin{equation}\label{eq:Shannon entropy}
H = -\sum_{i=0}^{n-1} p_i\log_{b}p_i,
\end{equation}
where $n$ denotes the number of gray levels (256 for 8-bit images), $b$ stands for the logarithmic base (returning bits when $b = 2$), and $p_i$ is the probability of a pixel having gray level $i$. However, although Shannon entropy considers compositional image information, it fails to account for spatial information, specifically the relationship between neighbouring pixels \cite{e20010019}.

Another entropy-based metric, the Gray Level Co-Occurrence Matrix (GLCM) entropy, unlike Shannon entropy, is a measure of how often pairs of pixel values occur in a grayscale image distribution \cite{4309314}. Taking into account this local spatial information, the GLCM is useful for various textural analysis tasks such as feature extraction for medical image segmentation \cite{9194196}. The GLCM entropy can be represented as
\begin{equation}\label{eq:GLCM}
H_g = -\sum_{i=0}^{n-1}\sum_{j=0}^{n-1}p_{(i, j)}\log_{b}p_{(i, j)},
\end{equation}
where $p_{(i, j)}$ is the probability of two pixels having gray levels $i$ and $j$ at a certain angle $\theta$ and distance $d$ away from each other. Despite GLCM better-capturing complexities within an image, it does not consider spatial patterns and global pixel relationships beyond its adjacent pairing.

\section{Methodology}
\label{sec:Methodology}
\subsection{Image Complexity Metric}
Our approach utilizes Larkin's delentropy, a metric identical to both the Shannon entropy and the GLCM entropy, but incorporating a new density function known as the \textit{deledensity} \cite{larkin2016reflectionsshannoninformationsearch}. By analyzing the relationship between the local and global features of an image, delentropy accounts for both an image's gradient vector field and pixel co-occurrence, encapsulating its spatial information as a whole. The deledensity, as a joint probability function, is formulated as
\begin{equation}\label{eq:Delentropy1}
\begin{aligned} {p_{(i,j)} = \frac{1}{4WH}\sum _{w=0}^{W-1}\sum _{h=0}^{H-1}\delta _{i,d_x(w,h)}\delta _{j,d_y(w,h)},} \end{aligned}
\end{equation}
where $d_{x}$ and $d_{y}$ denote the derivative kernels in the x and y direction, $\delta$ is the Kronecker delta to describe the binning operation required to generate a histogram, and $H$ and $W$ is the image's dimensions (height and width) \cite{khan2022leveraging}. By obtaining this, we can then calculate delentropy as
\begin{equation}\label{eq:Delentropy2}
\begin{aligned} {\text {DE} = - \frac{1}{2} \sum _{i=0}^{I-1}\sum _{j=0}^{J-1} p_{(i,j)} \log _b p_{(i,j)}, } \end{aligned}
\end{equation}
such that $I$ and $J$ represent the number of bins (discrete cells) in the 2D distribution, and the $\frac{1}{2}$ is derived from Papoulis' generalized sampling expansion \cite{1084284}.

To interpret this measure, yielding a high delentropy suggests an image has a high range of variation in pixel intensities and more sophisticated details. A low delentropy can be interpreted as a result of having a uniform distribution of pixel intensities, indicating simple structure and a less-detailed image.


Prior to any calculations, each image was preprocessed into an 8-bit, grayscale image. This ensured delentropy was calculated in a consistent, single-channel format throughout each dataset.

\subsection{GAN Selection}
Core to the experimental approach was the selection of two state-of-the-art GANs, SPADE-GAN \cite{park2019semanticimagesynthesisspatiallyadaptive} and StyleGAN 3 \cite{karras2020analyzingimprovingimagequality} on which to run the experimental pipeline. These networks have been widely adopted by the medical image synthesis community and empirically observed to produce superior-quality medical images when compared to predecessor GANs \cite{Skandarani2023}. StyleGAN 3's large community support and wide availability of its code repository along with its numerous configurations for different training settings were taken into account as well.

\subsection{GAN Pipeline}
For experiments, given our data-centric approach, StyleGAN 3 and SPADE-GAN were run with the official, publicly available implementations with default hyperparameters and no augmentations to each network’s architecture.

\paragraph{Preprocessing} We first set all images to a consistent 512x512 resolution. As such, training parameters were based on the size of the preprocessed images, as documented in the official implementations. SPADE-GAN additionally relies on segmentation masks to produce synthetic data. We used pre-existing annotations for ISIC-2018 and the Polyps Set. Because the Chest X-ray dataset did not have such annotations, masks were generated using TorchXRayVision \cite{Cohen2022xrv}. All experiments were performed on one NVIDIA A100 and three NVIDIA A40 GPUs. 

\paragraph{Training and Generation} The experimental pipeline was designed to identify the role of image dataset size in the image generation fidelity of selected GANs, for which to be compared to the image complexity distribution of each dataset. To that end, for each GAN training run, all parameters were held constant with the exception of the image set size, which was subsequently set to 500, 1000, and 2500 images, randomly sampled from the same dataset for each experimental run, respectively. The pipeline was designed to incorporate 500 images as a baseline for GANs to train with limited data. We facilitated experiments with 2500 images for a more comprehensive training run to better capture the underlying image distribution and use an intermediary set of 1000 images serving as a middle ground. For StyleGAN 3, all experimental runs were trained for 100 epochs; for SPADE-GAN, training iterated 50 epochs. The trained adversarial network was then used to generate synthetic images, the fidelity of which was then evaluated for each training set size. 

\paragraph{Evaluation} The Fréchet Inception Distance (FID) \cite{heusel2017gans} is a common metric used to evaluate the fidelity of the synthetically generated images for GANs \cite{BORJI2022103329}. Defined as the distance between the distributions of the ground truth and the generated images respectively, in our paper, we use the FID to assess the performance of the GAN (i.e. image fidelity) for each experimental run across both GANs. \textbf{A lower FID score signifies that a GAN is more proficient at generating synthetic data close to its target distribution}. From these data, we obtained fidelity curves for each dataset that describe how FID scores trend with increasing training set size.
\begin{figure}[htbp]
  \centering
  \includegraphics[width=\linewidth]{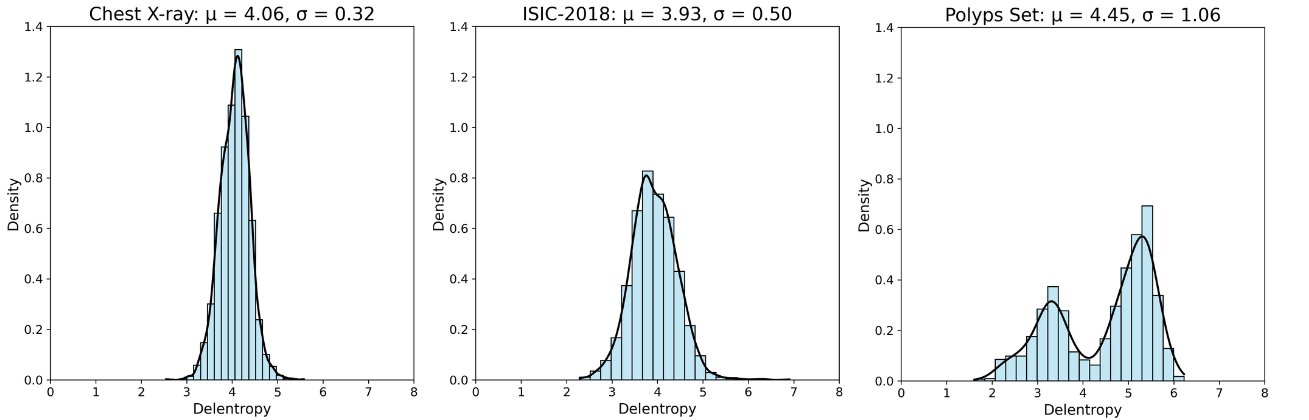}
  \caption{Delentropy distributions across each medical image dataset. A higher mean delentropy $\mu$ indicates a dataset with more complex images.}
  \label{fig:DelentropyComparisonGraph}
\end{figure}

\section{Experimental Results}
\label{sec:Experimental Results}
\paragraph{Datasets} We employed three medical image datasets: International Skin Imaging Collaboration 2018 Challenge (ISIC-2018) \cite{codella2018isic}, Chest X-Ray Images (Chest X-ray) \cite{kermany2018oct}, and Colonoscopy Polyp Detection and Classification (Polyps Set) \cite{DVN/FCBUOR_2021}. These datasets were chosen for their diversity in both perceptual complexity, ranging from relatively skin lesions to complex colon polyps, and imaging modality (dermoscopy vs. x-ray vs. colonoscopy). 

We carried out delentropy calculations as described in Section \ref{sec:Methodology} by using a publicly available implementation from Marchesoni \cite{stackexchange2023}. To effectively capture the overall complexity of each image dataset, we captured each dataset's delentropy distribution as displayed in Fig. \ref{fig:DelentropyComparisonGraph}.   

\begin{figure}
  \centering
  \includegraphics[width=\linewidth]{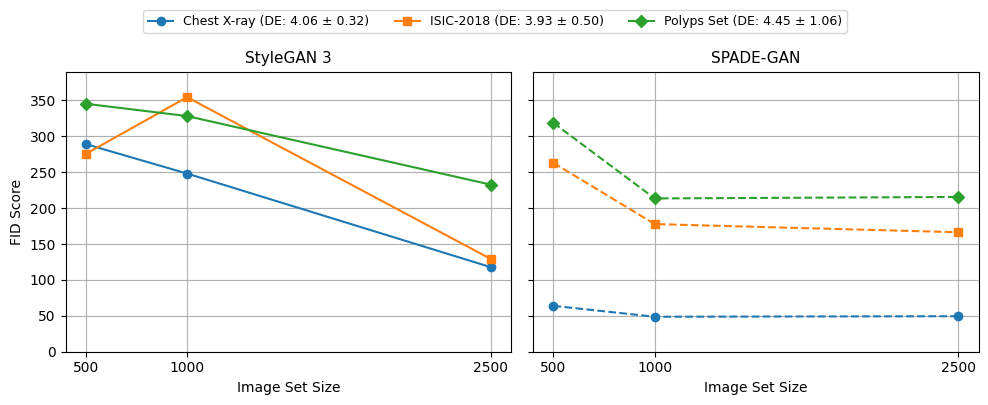}
  \caption{Fréchet Inception Distance (FID) curves comparing StyleGAN 3 and SPADE-GAN across each medical image dataset with varying sample sizes. Lower FID scores correspond to higher fidelity synthetic images.}
  \label{fig:FIDComaprisonGraph}
\end{figure}

Across the experimental runs, FID scores consistently decreased with increasing dataset size. On StyleGAN 3, synthesized images that had been generated by a GAN trained on 2500 images exhibited an average FID score reduction of 48\% when compared to those generated by a StyleGAN 3 that had been trained on a mere 500 images (Fig. \ref{fig:FIDComaprisonGraph}). SPADE-GAN experienced an analogous 31\% FID score reduction on average, though it is worth noting that FID reduction plateaued after only 1000 training images. 

Comparing both Fig. \ref{fig:DelentropyComparisonGraph} and Fig. \ref{fig:FIDComaprisonGraph}, one can see a general relationship between the delentropy distribution and the training performance of both GANs. As the spread of image complexities increases from a slender, peaked distribution to a broader, bimodal one, we see a corresponding increase in FID scores for each dataset sample size. The Chest X-ray dataset with the most homogeneous image complexities shown by a tall and narrow distribution, yields the lowest FID score after being trained for 2500 images, indicating that both GANs had easier training runs with this dataset. On the contrast, the Polyps Set—the dataset with the widest distribution and multiple complexity peaks—correlates with the highest FID scores for each dataset sample size, which suggests that the GAN was faced with a more challenging and unstable training run. Ultimately, this pattern shows a general inverse relationship—GAN performance decreases with an increasing spread of image complexities within a dataset.

\section{Discussion}
\label{sec:Discussion}
SPADE-GAN outperformed StyleGAN 3 across \textit{all datasets and training sizes}, with FID scores averaging 33\% lower, likely due to its architecture that incorporates segmentation masks for structural information, whereas StyleGAN 3 trained on raw image data alone, making it more difficult to generalize to high-delentropy datasets. Moreover, ISIC-2018 being an outlier can be attributed to its fluctuations in image complexity, reflected by the standard deviation in delentropy (Fig. \ref{fig:DelentropyComparisonGraph}). Despite having a lower mean delentropy, its spread likely resulted in difficulties in GAN training and learning the images' distribution, contrasting with the Chest X-ray dataset.

While the experimental results generally reflected an intuitive understanding of how image complexity and training data influence GAN training, the FID curves provide insightful details, offering a deeper perspective on these effects. SPADE-GAN exhibits both better quality results than StyleGAN 3 in the form of lower FID scores and more consistent training as evidenced by the smooth, non-overlapping FID curves (Fig. \ref{fig:FIDComaprisonGraph}). As aforementioned, performance plateaued after 1000 training images, suggesting that additional training data past that point may not help increase GAN performance as measured by FID score. This is also apparent in the generated images themselves, which exhibit little perceptual difference between those generated after 1000 training images and those generated after 2500 (Fig. \ref{fig:ImageComparisonGraph}). Contrast this with the StyleGAN 3 curves, which do not reach any noticeable plateau between 500 and 2500 training images. In fact, the increasingly negative slope values of the StyleGAN 3 graphs imply that StyleGAN 3 begins to better capture the images' features at a point past 1000 images, the exact whereabouts of which would need to be determined by a separate study.

The FID curves generated by this set of experiments set up a useful benchmark to which other potential training image data sets can be compared. For training sets that are of similar delentropy distributions and used to train StyleGAN 3 or SPADE-GAN, it is not unreasonable to predict that their training curves will be similar to those represented in Fig. \ref{fig:FIDComaprisonGraph}, though many more training set sizes and image sets are required before a truly comprehensive representation can be reached.

\paragraph{Broader Impacts} Our research on GANs for medical image synthesis may have positive and negative societal implications. On the positive side, it can enhance healthcare outcomes by improving the training of machine learning models with realistic synthetic data, therefore protecting patient privacy. Contrarily, potential negative impacts include the risk of maliciously generating fraudulent synthetic data and the possibility of reinforcing biases due to a lack of diversity of representing patient populations. These considerations demonstrate the importance of addressing both the benefits and potential risks associated with the use of GANs in the medical domain.


\section{Conclusion}
\label{sec:Conclusion}
In this work, we highlight the impact of image complexity on GAN performance in medical image synthesis. We empirically demonstrate a general inverse relationship: higher image complexity leads to poorer image fidelity results and lesser performance in GANs. Furthermore, we demonstrate FID curves showing healthcare professionals the possibility for the use of our benchmarks to gauge an estimate of data training requirements to achieve desirable results based on the image complexity distribution of a medical image dataset.

\section{Limitations}
\label{sec:Limitations}
Due to limited resources, experiments were only run on 500, 1000, and 2500 training images, leading to coarse-grained results. An extended study with a larger range and finer-grained increments would better elucidate exactly how FID scores respond to changes in training image dataset size. The use of FID scores as a sole evaluation metric also has its limitations, not necessarily correlating with human perceptual interpretations, something that is extremely important in the medical field where human doctors are still largely the source of truth. Skandarani \etal~\cite{Skandarani2023} shows that a lower FID score may not be a good measure of how well synthetic images can perform on a downstream task as well. Although this study specifically focused on GANs as the primary generative model, similar experiments extrapolated across other generative models such as stable diffusion may prove more relevant in the context of recent advancements in generative AI. More research with larger resources involving multiple evaluations of a similar experimental setup is required.

\section*{Acknowledgements}
We are grateful to Michael Lam and Kevin Zhu for their excellent mentorship, constructive feedback, and unwavering support throughout our research.

%
%
\bibliographystyle{splncs04}

\bibliography{main}



\end{document}